\documentstyle[aps,prbbib,epsfig,floats]{revtex} 
\begin{document} 
\begin{twocolumn}
\draft 
\wideabs{
\title{Proximity effects at ferromagnet-superconductor interfaces} 
\author{Klaus Halterman\cite{klaus} and Oriol T. Valls\cite{oriol}} 
\address{School of Physics and Astronomy and Minnesota Supercomputer 
Institute 
\\ University of Minnesota \\ 
Minneapolis, Minnesota 55455-0149} 
\date{\today} 
\maketitle 
\begin{abstract} 
We study proximity effects at ferromagnet superconductor 
interfaces by self-consistent numerical solution 
of the Bogoliubov-de Gennes equations for the continuum, 
without any 
approximations. Our procedures allow us to study systems
with long superconducting coherence lengths.
We obtain results for the pair potential, the 
pair amplitude, and the local density of states. We 
use these results to extract 
the relevant proximity lengths. We find that  the 
superconducting correlations in the ferromagnet exhibit 
a damped oscillatory behavior
that is reflected in both the pair amplitude and
the local density of states.
The characteristic length scale
of these oscillations
is approximately inversely proportional to the exchange field,
and is  independent of the superconducting coherence length
in the range studied. 
We find the superconducting coherence 
length
to be nearly independent of the 
ferromagnetic polarization.
\end{abstract} 
 }
% body of paper here 
\section{Introduction} 

In recent years,  technological advances
in materials growth and  fabrication techniques
have made it possible to create heterostructures including
high quality ferromagnet/superconductor (F/S) interfaces. 
These systems have great intrinsic scientific importance and
potential device applications,
including 
quantum computers and magnetic information storage\cite{blatter,oh,tagirov}.
This has led
to renewed interest in proximity effects involving magnetic
and superconducting compounds.
Understanding how  proximity effects
modify electronic properties 
near F/S interfaces
is constantly becoming
more important as
the rapid growth of nanofabrication technology continues.

The  juxtaposition of a ferromagnet and a superconductor
can result\cite{parks} in a
spatial variation of  magnetic and superconducting correlations
in both materials.
The leakage of superconducting correlations
into the non-superconducting material is an example of
the superconducting proximity effect.
Similarly, the spin polarization may extend into the superconductor
and modify its properties, creating a magnetic proximity effect.

In general, 
if one is interested in 
a microscopic solution  of the F/S proximity
effect problem valid at all length scales,
one must solve the appropriate equations, e.g. the Gor'kov\cite{fw},
or Bogoliubov-de Gennes\cite{bdg} (BdG) equations in a self-consistent
manner and with as few approximations as possible.
In practice, approximations are often made in the basic
equations. Further, in many cases a simple form for  the
pair potential $\Delta({\bf r})$ is
assumed, usually  a constant in the
superconductor region, and zero elsewhere
is  used.
Such crude non self-consistent
treatments 
have been widely applied because
of their simplicity. 
However they are valid
typically  
only for length scales much longer than the superconducting
coherence length, which characterizes 
the depletion of the pair potential in
the superconductor near the interface,
or in the case where the non-superconductor is 
very thin.\cite{arnold}
The superconducting proximity effect  is linked to
the  phenomenon of Andreev reflection\cite{andreev}.
This is the process where at the interface,
an electron is reflected as a hole, transmitting
a Cooper pair into the superconductor and vice versa.
The  inhomogeneity in $\Delta({\bf r})$
creates a potential well
for quasiparticles, causing
electon-hole scattering, and 
subsequent bound states
below the maximum value of  $\Delta({\bf r})$.

There are several quantities that can be studied, theoretically
or experimentally, in the
context of characterizing proximity effects.
The traditional description\cite{parks} of
the superconducting proximity effect is through a characteristic
proximity length which can be associated with the behavior of the 
{\it pair amplitude}, $F({\bf r})$,
the probability amplitude to find a Cooper pair
at point ${\bf r}$.  This quantity does not
vanish identically inside the non-superconductor.
This is in contrast
to the pair potential $\Delta({\bf r})$, which
is of
limited use, since it is zero inside the non-superconducting
material unless it is arbitrarily assumed that a small nonvanishing
pairing interaction exists there. 
An additional important quantity, 
which is now
experimentally accessible thanks to improved
STM technology\cite{moussy} which allow local spectroscopy to be performed,
is the local density of states (DOS). This quantity
reflects the one-particle
energy spectrum, and therefore
one aspect of
the proximity effect.

For a nonmagnetic normal
metal in contact with a superconductor,
the proximity effect
has been much studied and well understood
for many years.\cite{parks}
For clean systems, if the non
self-consistent step function 
for the pair potential is used,
solutions to the microscopic equations are relatively easy to 
obtain.\cite{zaitlin,degennes2,mcmillan,wohlman}
Other  
approaches 
involve first simplifying the basic equations.
One can, for example,  integrate out
the energy variable in the Gor'kov equations.
The resultant (quasiclassical) Eilenberger\cite{eilen} equations have the
advantage of being first order,
and therefore easier to solve.
They can be extended to
systems of arbitrary impurity concentration.\cite{pilgram}
Results 
that calculate the pair potential
self-consistently are more sparse.
The  Eilenberger equations have been solved numerically,
\cite{kieselmann} and 
the DOS was calculated, with
comparisons made between self-consistent and non-self-consistent 
results.
For systems in which
the electron mean free path
is much shorter than the superconducting coherence
length, when  the Eilenberger equations
can be reduced to the simpler Usadel equations\cite{usadel},
a calculation of the DOS\cite{belzig} has been performed.
Numerical approaches which do not require
simplifying the starting equations
are  possible, although rare.
Numerical self consistent solutions of  the full 
Gor'kov equations
in heterogeneous systems have  been obtained,\cite{branko,branko2} 
and from these the density of states and pair potential
of normal metal-superconductor bilayers 
and multilayers were calculated.\cite{branko2}

When the normal metal is replaced by a ferromagnet,
the theoretical situation is much less satisfactory.
The presence of the exchange field in the ferromagnet
makes the overall physical and mathematical picture
of the proximity effect in F/S systems
quite different from its non-magnetic
counterpart.
Since on the magnetic side Fermi surface quasiparticles
with different spins have different wavevectors, numerical
solution becomes much more difficult, as
matrices in wavevector space become more complicated,
and approximate diagonalization methods such as those employed in
Refs.~(\onlinecite{branko,branko2}) 
cannot be used. 
The only   existing microscopic
numerical self-consistent
calculations\cite{ting,kuboki} addressing
the proximity effect at an F/S interface, 
are based on
an extended Hubbard model in real space. This
is adequate only when the coherence length
is very short, and the material parameters
used\cite{ting} were unrealistic.
Analytic
work is similarly hampered. The traditional\cite{parks}
way out is to conjecture a dependence of the proximity length
on the exchange field, but the underlying
assumption, while plausible, has never been proved and   has been
recently labeled\cite{lg} as being just {\it ad hoc}.
Physically, the spin imbalance
in F results in a modified  Andreev process,
since the electron and hole occupy
opposite spin bands.\cite{beenakker}
The exchange field causes the quasiparticles comprising a
singlet Cooper pair to have different wavevectors,
so that
the pair amplitude
in the ferromagnet
becomes spatially modulated.\cite{demler}
Such oscillations were first investigated  long
ago by  Fulde and Ferrell\cite{fulde}
and  Larkin and Ovchinnikov.\cite{larkin}
The resulting oscillations in $F({\bf r})$
induce 
oscillations (about
the normal state value)
in the local
density of states (DOS)
as a function
of distance from the interface.
These oscillations
have been studied theoretically\cite{zareyan} 
(but non  self-consistently) by using the
Eilenberger equations,
and good agreement was found with experiment.\cite{kontos}
The  Usadel
equations revealed similar behavior.\cite{buzdin}
Quasiclassical approaches
are clearly no substitute for a microscopic analysis 
which is required to study the case when the
correlation length is not small.

Thus, the theoretical situation for
the F/S interface problem is unsatisfactory. In this paper, 
we attack this problem by obtaining
numerical, fully self-consistent solutions
for the continuum  BdG equations
for a ferromagnet
in contact with an  $s$-wave  superconductor.
Our numerical iterative methods overcome the
technical difficulties associated
with the  different Fermi wavevectors, alluded to above,
and allow us to focus on the case of longer
superconducting coherence lengths
in the clean limit.  We
are able also to allow for 
different bandwidths in the two materials 
(Fermi wavevector mismatch\cite{zv}).
The full BdG equations  that are our starting point provide  a
rigorous, microscopic method
for studying inhomogeneous
superconductors and their interfacial
properties, and
have the advantage that their solution
provides
the quasiparticle amplitudes and excitation 
energies.
The
resulting wave functions and energies
are  used to
compute physically relevant quantities.
We extract the relevant lengths from analysis
of $F(\bf r)$ and  investigate
the local DOS as a function of position
on both sides of the F/S interface.
Our results put the entire theory of the F/S proximity
effect on firmer grounds, confirm some of the features
previously obtained approximately, and uncover new ones.

This paper is organized as follows.
In Sec. \ref{methods}, we
introduce the spin-dependent
BdG equations, and the methods we
employ to
extract the pair potential, 
the pair amplitude, and the local DOS. In Sec. \ref{results} we discuss the
physical parameters we will use and present the results.
Finally in
Sec. \ref{conclusions}, we summarize the results
and discuss future work.

\section{Method} 
\label{methods}
In this section we present the basic equations we use
for a system containing 
a ferromagnet/superconductor (F/S) interface and the methods 
we employ for their self-consistent solution. 
After self-consistency for the pair potential is achieved,
we can then calculate other
physically relevant quantities such as the 
pair condensation amplitude and the local DOS.

The system we consider is  semi-infinite 
and uniform in the $x,y$ directions and 
confined to the region $0<z<d$, with the 
F/S interface located at 
$z=d'$ and the superconductor 
in the 
region $z>d'$.
We will take here $d$ and $d'$ larger than the other relevant lengths in 
the problem 
in order to study the interface between two bulk materials. 
 
We begin with a brief review of the starting equations in order 
to clarify our notation and conventions, including spin 
and choice of parameters. 
For a spatially inhomogeneous system, a complete description of the 
quasiparticle 
excitation 
spectrum along with the quasiparticle amplitudes is given by 
the BdG equations\cite{bdg}.  
In the absence of an applied magnetic  
field, the system is described, using the usual second quantized form,
by an effective mean field 
Hamiltonian, 
\begin{eqnarray} 
\label{heff} 
&&{\cal{H}}_{\rm eff}=\sum_{\sigma,\sigma'}{\int{d^3r\Bigl\lbrace  
\widehat{\psi}^\dagger_\sigma ({\bf r})  
{\cal H}_{0}({\bf r})\widehat{\psi}_\sigma({\bf r}) + 
\widehat{\psi}^\dagger_\sigma({\bf r})  
{h}_{\sigma \sigma'}({\bf r})\widehat{\psi}_{\sigma'}({\bf r})}}  \nonumber \\
&&+\frac{1}{2}\eta_{\sigma \sigma'}[\Delta({\bf r})  
\widehat{\psi}^\dagger_\sigma({\bf r})\widehat{\psi}^\dagger_{\sigma'}({\bf r})  
+\Delta^*({\bf r})   
\widehat{\psi}_\sigma({\bf r})\widehat{\psi}_{\sigma'}({\bf r})]\Bigr\rbrace,  
\end{eqnarray}  
where  
$\Delta({\bf r})$ is the pair potential, to be 
calculated self-consistently, greek indices denote spin,
$\widehat{\eta}=i\widehat{\sigma}_y$ (the $\widehat{\sigma}$'s are the usual Pauli 
matrices),
and $\widehat{h}({\bf r}) = -h_0 \widehat{\sigma}_z\Theta(d'-z)$ is the 
magnetic exchange  
matrix. The step function in this term reflects the assumption that 
the exchange field arises from the electronic structure in the  F 
side. 
The single-particle Hamiltonian is given by, 
\begin{equation} 
\label{hzero}
{\cal H}_{0}({\bf r}) =-\frac{1}{2m}\nabla^2+U_0({\bf r})-\mu.  
\end{equation} 
Here $\mu$ is the chemical potential, 
$U_0$  
is the spin 
independent mean field  term,
and we have set $\hbar = k_B =1$.
 
The BdG equations are derived by setting up the
diagonalization of the
effective Hamiltonian
via a Bogoliubov 
transformation, which in our notation is written as 
\begin{mathletters}
\label{trans}
\begin{eqnarray} 
\label{trans1}
\widehat{\psi}_\uparrow({\bf r})&=& 
\sum_n{[ u_{n \uparrow}({\bf r})\widehat{\gamma}_{n}- 
{ v^*_{n \uparrow}({\bf r})\widehat{\gamma}^{\dagger}_{n}]}}, \\
\widehat{\psi}_\downarrow({\bf r})&=& 
\sum_n{[ u_{n \downarrow}({\bf r})\widehat{\gamma}_{n}+ 
{ v^*_{n \downarrow}({\bf r})\widehat{\gamma}^{\dagger}_{n}]}},  
\label{trans2} 
\end{eqnarray} 
\end{mathletters}
where 
$\widehat{\gamma}$ and $\widehat{\gamma}^{\dagger}$ are  Bogoliubov 
quasiparticle annihilation and creation operators 
respectively, and 
$n$ 
labels the  relevant quantum 
numbers.  
The quasiparticle amplitudes $u_{n \alpha}$ and $v_{n \alpha}$ are 
to be determined 
by requiring that Eqs.(\ref{trans}) 
diagonalize Eq.(\ref{heff}). 
The resulting\cite{bdg} BdG equations  read, 
\begin{mathletters}
\label{set} 
\begin{eqnarray} 
\label{set1} 
({\cal H}_0+h_{\sigma\sigma}) u_{n \sigma}+ 
\sum_{\sigma'}{ \rho_{\sigma \sigma'} 
\Delta({\bf r}) v_{n \sigma'}({\bf r})} &=&  
\epsilon_n u_{n \sigma}({\bf r}), \nonumber \\  \\
-({\cal H}_0 + h_{\sigma\sigma}) v_{n \sigma}+  
\sum_{\sigma'}{ \rho_{\sigma \sigma'} 
\Delta^*({\bf r}) u_{n \sigma'}({\bf r})} &=& 
\epsilon_n v_{n \sigma}({\bf r}), \nonumber \\
\label{set2} 
\end{eqnarray} 
\end{mathletters} 
where $\widehat{\rho}\equiv\widehat{\sigma}_x$, and the $\epsilon_n$ are the 
quasiparticle energy eigenvalues
measured with respect to the chemical potential.
Equations (\ref{set}) 
must be supplemented by the  
self consistency condition for the pair potential, 
$\Delta({\bf r})=g({\bf r})  
\langle \widehat{\psi}_{\uparrow}({\bf r}) 
\widehat{\psi}_{\downarrow}({\bf r}) \rangle$,  
which in terms of the 
quasiparticle amplitudes reads, 
\begin{equation}  
\label{del2} 
\Delta({\bf r}) =\frac{g({\bf r})}{2} 
\sum_{\sigma,\sigma'}{\rho_{\sigma \sigma'}}{\sum_{n}}'{ 
u_{n\sigma}({\bf r})v^{*}_{n \sigma'}({\bf r})\tanh(\epsilon_n/2T)}, 
\end{equation} 
where
$g({\bf r})$ is the effective superconducting coupling. 
We take this quantity 
to be a constant in the superconductor, and to vanish 
outside of it. This is analogous to the assumption made
for $\widehat{h}$. Our method does not require
that a small nonzero value of $g$ be assumed
in the non-superconducting side.
The prime on the sum in (\ref{del2}) reflects 
that the sum is only over 
eigenstates  
with 
$|\epsilon_n| \leq \omega_D$, where $\omega_D$ is the cutoff (Debye) energy. 
The normalization condition for the quasiparticle amplitudes 
in our geometry
is,
\begin{equation}
\label{norm}
\sum_{\sigma} \int^{d}_{0}{d^3r  
[|u_{n \sigma}({\bf r})|^2+ |v_{n \sigma}({\bf r})|^2]} = 1. 
\end{equation}
 
The Hamiltonian is translationally invariant in any plane parallel to 
the interface, 
therefore the component of the wavevector 
perpendicular to the $z$ direction, $k_{\perp}$, is a good quantum number. 
We can then write 
\begin{mathletters} 
\begin{eqnarray} 
u_{n\sigma}({\bf r}) = u^{\sigma}_{n}(z) 
e^{i {\bf k}_{\perp} \cdot {\bf r}}, \\ 
v_{n\sigma}({\bf r}) = v^{\sigma}_{n}(z) 
e^{i {\bf k}_{\perp} \cdot {\bf r}}, 
\end{eqnarray} 
\end{mathletters} 
where ${\bf k}_{\perp} = (k_x,k_y,0)$.  
Eqs.(\ref{set})   
then become one 
dimensional 
BdG equations, 
\begin{mathletters}
\label{newset} 
\begin{eqnarray} 
\label{newset1} 
\Bigl[-\frac{1}{2m}\frac{\partial^2}{\partial z^2} 
+\varepsilon_{\perp} 
+h_{\sigma \sigma}(z)-{\mu}\Bigr] 
u^{\sigma}_n(z) \nonumber \\
+\sum_{\sigma'}{\rho_{\sigma \sigma'} 
\Delta(z) v^{{\sigma}}_n(z)} = \epsilon_n u^{\sigma}_n(z), \\ 
-\Bigl[ -\frac{1}{2m}\frac{\partial^2}{\partial z^2} 
+\varepsilon_{\perp} 
+h_{\sigma \sigma}(z)-{\mu}\Bigr] 
v^{{\sigma}}_n(z) \nonumber \\
+\sum_{\sigma'}{\rho_{\sigma \sigma'} \Delta(z) u^{\sigma}_n(z)}=  
\epsilon_n v^{{\sigma}}_n(z), 
\label{newset2} 
\end{eqnarray} 
\end{mathletters} 
where $\varepsilon_{\perp}$ is the transverse kinetic 
energy, and
we have absorbed 
the mean field term by 
a shift in the zero of the energies. 
One can assume $\Delta(z)$ to be real without loss of generality. 

We can now solve Eq.~(\ref{newset})
by expanding the quasiparticle  
amplitudes in terms 
of a complete set of functions $\phi_m(z)$, 
\begin{mathletters} 
\begin{eqnarray} 
u^{\sigma}_n(z)&=&\sum_{m=1}^{N} u^{\sigma}_{n m} 
\phi_m(z),\\  
v^{\sigma}_n(z)&=&\sum_{m=1}^{N} v^{\sigma}_{n m} 
\phi_m(z).  
\end{eqnarray} 
\end{mathletters} 
A set of functions appropriate for  
our setup  and geometry is that of the 
normalized free particle wavefunctions of a one-dimensional 
box, 
\begin{equation} 
\phi_m(z) = \sqrt{\frac{2}{d}}\sin(k_m z),\qquad k_m = \frac{m\pi}{d}. 
\end{equation} 
If there was only one Fermi wavevector in the problem,
the upper limit in the sum, $N$, would be determined by
that wavevector and $\omega_D$ in the usual way.\cite{branko}
But since this is not the case some care is required. 
The appropriate cutoff for this problem is given by
\begin{equation} 
N = [(k_{FX} d/\pi)\sqrt{1+\omega_D/\mu}], 
\label{number}
\end{equation} 
where $k_{FX}$  is the largest Fermi wavevector in either the S or F side
(see below) and the brackets denote the integer
value of the expression they enclose.
In a similar way, we can also expand the pair potential, 
\begin{equation}
\label{deltaexp} 
\Delta(z) = \sum_{q=1}^{N} \Delta_q \phi_q (z). 
\end{equation} 
After inserting these expansions into Eqs.(\ref{newset})
and making use of the orthogonality 
of the chosen basis, we obtain
the following equations for the  
the matrix elements, 
\begin{mathletters} 
\label{meelements}
\begin{eqnarray} 
\label{matrixelements1} 
&&\Bigl[\frac{k^2_{q}}{2m}+ \varepsilon_{\perp}\Bigr] 
u^{\sigma}_{n q} - 
\sum_{q'}{\Bigl[(E_{FM} - h_{\sigma \sigma})F_{q q'}+  
E_{FS}S_{q q'}\Bigr] 
u^{\sigma}_{n q'}} \nonumber \\
&&+\sum_{\sigma'} 
{\sum_{p,p'}}{\rho_{\sigma \sigma'}\Delta_p J_{p p' q} 
\,v^{{\sigma}}_{n p'}}=\epsilon_n 
u^{\sigma}_{n q}\\ 
&&-\Bigl[\frac{k^2_{q}}{2m}+ \varepsilon_{\perp}\Bigr] 
v^{{\sigma}}_{n q}+ 
\sum_{q'}{\Bigl[(E_{FM} - h_{\sigma \sigma})F_{q q'} +  
E_{FS}S_{q q'}\Bigr] 
v^{{\sigma}}_{n q'}} \nonumber \\
&&+\sum_{\sigma'}{\sum_{p,p'}}{\rho_{\sigma \sigma'} 
\Delta_p J_{p p' q} 
\,u^{\sigma}_{n p'}}=\epsilon_n 
v^{{\sigma}}_{n q}. 
\label{matrixelements2} 
\end{eqnarray} 
\end{mathletters}
In writing each term in Eq. (\ref{meelements})
we have  taken care to measure the chemical potential from the same
origin (bottom of the same band) as the corresponding energies.
Because of the magnetic polarization and  
possible differences in carrier densities 
between the ferromagnet 
and superconductor,
there are up to three different Fermi wavevectors involved
in the problem,
the two corresponding
to spin up
and and spin down on the F side,
and one in the superconductor.  On the F side we have introduced
$E_{FM}$ through ${k^{2}_{F\uparrow}}/{2m} 
\equiv E_{F \uparrow} \equiv E_{FM}+h_0$,
${k^{2}_{F\downarrow}}/{2m} \equiv E_{F \downarrow} \equiv E_{FM}-h_0$.
On the S side, we have 
${k^{2}_{FS}}/{2m}= E_{FS}$, where $E_{FS}$
is the appropriate bandwidth. It has 
been shown,\cite{zv} that  
Fermi wavevector mismatch in
F/S tunneling 
junction spectroscopy leads to 
nontrivial differences in the conductance spectrum. 
The matrix elements in (\ref{meelements}) are given by, 
\begin{mathletters} 
\begin{eqnarray} 
F_{q q'}&=&C_{q'-q}(d')-C_{q'+q}(d'), \\ 
S_{q q'}&=&\delta_{q q'}- F_{q q'}, \\ 
J_{p p' q}&=&-\frac{1}{\sqrt{2 d}}\Bigl[E_{q+p'-p}(d)-E_{q+p'-p}(0) 
          +E_{p+p'-q}(d) \nonumber \\
&-&E_{p+p'-q}(0)  
          + E_{p+q-p'}(d)-E_{p+q-p'}(0)  \nonumber \\
           &-&E_{p+p'+q}(d)+E_{p+p'+q}(0)\Bigr], 
\end{eqnarray} 
\end{mathletters} 
where we have defined 
$C_m(z) \equiv \sin(k_m z)/(\pi m)$, $E_m(z) \equiv  
\cos(k_m z)/(\pi m)$, for $m\neq0$, and $E_0(z) \equiv 1$.
The self-consistency condition now reads, 
\begin{equation} 
\label{selfcon} 
\Delta_{q} = \frac{g}{2} 
\sum_{p,p'}{K_{p p' q} 
\sum_{\sigma,\sigma'}{{\sum_{n}}'{}\rho_{\sigma \sigma'} 
{u^{\sigma}_{n p}v^{{\sigma'}}_{n p'}} 
\tanh(\epsilon_n/2T)}}, 
\end{equation} 
where
the quantum numbers $n$ include $\varepsilon_{\perp}$ and  
a longitudinal index $m$, the sum being limited 
by the restriction mentioned below Eq.(\ref{del2}), and we have, 
\begin{eqnarray} 
K_{p p' q}& =&-\frac{1}{\sqrt{2 d}}\Bigl[E_{q+p'-p}(d)-E_{q+p'-p}(d') 
          +E_{p+p'-q}(d) \nonumber \\
&-&E_{p+p'-q}(d')  
          + E_{p+q-p'}(d)-E_{p+q-p'}(d')  \nonumber \\
           &-&E_{p+p'+q}(d)+E_{p+p'+q}(d')\Bigr]. 
\end{eqnarray} 
Finally, the normalization condition, Eq.(\ref{norm}),
in terms of the expansion coefficients, is 
\begin{equation}
\sum_{\sigma} \sum_{m}[|u^{\sigma}_{n m}|^2+ |v^{\sigma}_{n m}|^2] = 1. 
\end{equation} 

It is very difficult to
solve Eqs.(\ref{meelements})  
numerically as they stand, for
large sizes. The required effort can 
be considerably reduced by 
solving   
for $u_{n q}^{\uparrow},v_{n q}^{\downarrow}$ only, allowing for both 
positive and negative energies. The solutions for 
$u_{n q}^{\downarrow},v_{n q}^{\uparrow}$  
are then obtained  
via the transformation: 
$u_{n q}^{\uparrow} \rightarrow v_{n q}^{\uparrow},  
v_{n q}^{\downarrow} \rightarrow -u_{n q}^{\downarrow},  
\epsilon_n \rightarrow -\epsilon_n.$
This simplification follows from the form of the exchange
matrix, below  Eq.(\ref{heff}).
Formally, the exchange field breaks the
rotational invariance in spin space,\cite{radovic}
however, there are no spin flip effects, 
so that the four equations 
(\ref{meelements}) 
split into  two equivalent sets of equations.
 
For any fixed $\varepsilon_{\perp}$ we can now cast 
Eqs. (\ref{meelements}) 
as a
$2N\times2N$ matrix eigenvalue problem, 
\begin{eqnarray} 
\label{nset1} 
\left[  
\begin{array}{cc} 
H^{+} & D \\  \\
D & H^{-} 
\end{array} 
\right] 
\Psi_n 
= 
\epsilon_n 
\,\Psi_n, 
\end{eqnarray} 
where $\Psi_n$ is the column
vector corresponding to $\Psi_n^T = 
(u^{\uparrow}_{n1},\ldots,u^{\uparrow}_{nN},v^{\downarrow}_{n1}, 
\ldots,v^{\downarrow}_{nN}).$ 
The matrix elements are 
\begin{mathletters}
\label{basic} 
\begin{eqnarray} 
H^{+}_{q q'}&=& [\frac{k^2_q}{2m} + \varepsilon_{\perp}]  
\delta_{q q'} - E_{F \uparrow} F_{q q'}-E_{F S} S_{q q'},\\ 
H^{-}_{q q'}&=&-[\frac{k^2_q}{2m}+ \varepsilon_{\perp}] 
\delta_{q q'} 
            +E_{F \downarrow} F_{q q'} + E_{FS} S_{q q'}, \\ 
D_{q q'}&=&\sum_{p}{\Delta_p J_{p q q'}}.
\end{eqnarray} 
\end{mathletters} 

The basic method of self consistent solution 
of  Eqs. (\ref{nset1}) and (\ref{selfcon}) works  as follows: 
we first choose an initial trial form 
for the  $\Delta_p$. 
We then  find, by numerical diagonalization, 
all the eigenvectors and  eigenvalues 
of the matrix 
in Eq.(\ref{nset1}), for every value of $\varepsilon_{\perp}$
consistent with the energy cutoff (see Eq. (\ref{number})).
The formally continuous variable $\varepsilon_{\perp}$
is discretized for numerical purposes.
The calculated eigenvectors 
and eigenvalues are then summed according to 
Eq.(\ref{selfcon}), and a new pair 
potential  is 
found.
This new pair potential is then  substituted\cite{conver} 
into the entire set of eigenvalue equations, and a new set of eigenvalues 
and eigenvectors is obtained, from which in turn a new pair potential 
is constructed. 
The whole process is repeated 
until convergence is obtained, that is, until  
the maximum relative change in the pair potential 
between successive iterations is 
sufficiently small (see below). 
As an initial guess for the pair potential 
one can use, in the first instance,   
a step function of the  bulk value, $\Delta_0$, in  
the superconductor. The initial
$\Delta_p$ are then obtained by inverting (\ref{deltaexp}).
After self-consistent 
results for $\Delta_p$ for one set of parameter values 
have been obtained,
those results can be used as 
the initial guess for a case 
involving a nearby set of parameter values. This process
reduces the number of required iterations considerably. 
The final self-consistent result 
is insensitive to the initial choice.
By using these methods, 
it is then possible, as we shall 
see, to obtain results even when the coherence length is long. 
 
This general procedure immediately yields the self 
consistent results for the pair potential. 
As mentioned in the introduction,
this quantity gives valuable information 
regarding superconducting correlations on the S side ${\it only}$, 
since it vanishes on the F side 
where $g({\bf r})=0$. 
Insight into the superconducting correlations 
on the F side, and the extraction of the proximity effect 
in the ferromagnet, is most easily obtained 
by considering \cite{parks,degennes} the {pair amplitude}, 
\begin{equation} 
F({\bf r}) =\Delta({\bf r})/g({\bf r}),  
\label{ca} 
\end{equation} 
which has a finite value on both sides of the interface. 
One can also study proximity effects through 
another quantity which   
is directly related to observation. This is 
the local density of states (DOS), given by\cite{gygi}
\begin{equation} 
\label{dos}
{N}(z,\epsilon) 
=\sum_{\sigma}{}{\sum_{n}}' 
{[u^2_{n\sigma}(z)\delta(\epsilon-\epsilon_n) 
+v^2_{n\sigma}(z)\delta(\epsilon+\epsilon_n)]}. \\ 
\end{equation} 

In the next section we will first consider the relevant set 
of dimensionless 
parameters in the problem, and how we implement the general 
procedures discussed above for a wide range of the values of these 
parameters. We then 
discuss our results, and  investigate 
the length scales relevant to 
the variation of the pair potential, the pair amplitude and the DOS. 

\section{Results} 
\label{results}

Before discussing our numerical techniques
and results
for the model outlined in the previous section,
we have to introduce a
convenient set of  dimensionless parameters for the problem.
First, there are two dimensionless ratios 
arising from the three material parameters $E_{FS}$, $E_{FM}$ and $h_0$.
We choose the ratio $I \equiv h_0/E_{FM}$ as the 
dimensionless exchange field parameter we
will vary to study different degrees of polarization for the F side.
$I$ varies between $I$ = 0 
when one has a normal (nonmagnetic) metal and $I$ = 1, the half-metallic
limit. In this work we choose the second ratio so that 
$E_{F\uparrow}/E_{FS}$ = 1 at the value of $I$
under consideration.  
Next, we have to consider
the superconducting parameters.  
We have chosen to present here results for
$T=0$, postponing the study of temperature effects for future work. 
We then need to specify 
the dimensionless Debye frequency $\omega \equiv \omega_D/E_{FS}$ and 
the dimensionless length scale 
$k_{FS} \xi_0$, where $\xi_0$ is the usual zero-temperature  coherence 
length related to
other quantities
by the BCS relation, $k_{FS} \xi_0 = (2/\pi)(E_{FS}/\Delta_0)$.
Throughout, we will keep  
the relatively unimportant parameter $\omega$ fixed at $0.1$,
and  present results for two different 
values of $k_{FS} \xi_0$, $50$ and $200$.
Thus, our method can handle coherence lengths two orders of
magnitude larger than what has been achieved through the use\cite{ting}
of tight binding methods. 

We also have to consider the purely
computational parameters. 
These are determined by the overall size of the system, measured in terms of
$k_{FS} d$, and  the ratio $d'/d$. 
Our two choices of $\xi_0$ demand
different system sizes, since
the length scale over which the pair potential
reaches its
bulk value in S is determined (see below) by $\xi_0$.
Thus, we need $d\gg\xi_0$ in order to study an interface between bulk systems.
Thus,
we take $k_{FS} d = 1000$, $d'/d = 600/1000$,  for
$k_{FS}\xi_0=50$,  
and $k_{FS} d = 1700$, $d'/d = 750/1700$
for  $k_{FS}\xi_0=200$.
These slab widths  allow
us to investigate fully the bulk
proximity effects that occur on
both sides of the interface.

The  computational work required is chiefly
determined by the system size.
As outlined in the previous section,
we must numerically diagonalize the Hamiltonian matrix and 
calculate the eigenenergies and eigenvectors
for each $\varepsilon_{\perp}$.
Each value of  $\varepsilon_{\perp}$ requires diagonalizing
a matrix of size 
$2N\times 2N$, where $N$ is defined in Eq. (\ref{number}).
For the large values of $d$ required by our assumed values of $\xi_0$
this matrix size exceeds 1100. 
The number of discretized transverse energies, $N_{\perp}$,
must be chosen large enough so that the results are not affected
by it. 
The required value depends on the quantity being
studied. For $\Delta(z)$, and $F(z)$, a value  of
$N_{\perp}=500$ was found
to be sufficient even for the longer coherence length. For
the local DOS, we used $N_{\perp}=1000$ in both cases.
These diagonalizations must be performed at each step in 
the iteration process described below Eq.(\ref{basic}). 
The basic diagonalization process employed
a procedure whereby
the symmetric matrix, Eq.(\ref{nset1}), is transformed
into tridiagonal
form, and then the eigenvalues and eigenvectors
are computed by the QL\cite{essl} algorithm.
The iteration process was concluded when
the maximum relative error between successive iterations
of the pair potential at any point
was less than $10^{-3}$.
A smaller relative error would
require more computation time,
but we verified that no appreciable difference in the results
ensued. A number of checks were performed, 
including reproducing the correct
wavefunctions and energies for 
the limiting case of 
a single semi-infinite superconductor,
ferromagnet or normal metal, and
also verifying that
in the limit of an entirely superconducting sample the correct
finite size oscillations\cite{tb,falk} of the pair potential
were obtained, with the correct $\xi_0$ dependence.\cite{branko}

\subsection{Pair potential}
We begin by presenting in Fig.\ref{pp1}
our   self consistent results for the pair potential, $\Delta(z)$ 
(normalized to the bulk value $\Delta_0$), which we plot
as a function of the dimensionless variable $Z \equiv k_{FS}z$.
In the four panels of the left column we show results   
for $k_{FS}\xi_0=50$ for four evenly spaced values of $I$ 
ranging from zero to unity. In
the corresponding panels in the right column we have
results 
\begin{figure}[t]
{\epsfig{figure=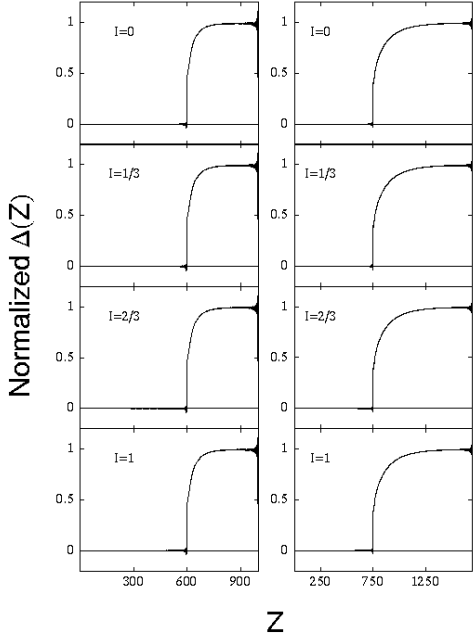,width=.45\textwidth}}
\newline
\caption{The self consistent pair potential  
$\Delta(z)$, normalized to the bulk value $\Delta_0$, 
is plotted 
as a function of dimensionless distance $Z \equiv k_{FS} z$.
The left column is for  $k_{F}\xi_0=50$, with 
the F/S interface  at $Z=600$. The right column
is for  $k_{F}\xi_0=200$, the interface is then at $Z=750$.
In both cases results for the same 
four  exchange  
fields (indicated by the labels) are shown.}
\label{pp1}
\end{figure} 
for  $k_{FS}\xi_0=200$ at the same values of the exchange field.
The pair potential always vanishes on the F
side since
we assumed $g({\bf r})=0$ in that region.
All of the panels show that
on the S side,
the normalized pair potential  
rises near the interface 
and then eventually reaches its bulk value over
a length scale determined by the coherence length 
$\xi_0$.
Comparing the top panels
in each column, where $I=0$, with the others in the figure,
where the exchange field can be large, 
we see that for all four values of $I$ and a given value
of $\xi_0$, the 
characteristic depletion near
the interface is nearly independent of $I$.
It can be concluded
therefore, that the magnitude of the exchange field
has little  
effect on $\Delta(z)$ and that the effective coherence length
in the superconducting side of 
the F/S interface is only an extremely weak function
of the strength of the ferromagnetic exchange field.
Similar findings were obtained in Ref.\onlinecite{ting},
in the short $\xi_0$ limit.
The general shape of the curves is the same
for both values of $\xi_0$, indicating that the effect of this quantity
is merely a rescaling of the relevant length
which governs the interface depletion. 
Near the surface-vacuum  boundary in S,
the pair potential exhibits  atomic scale
(a  distance of order $1/k_{FS}$)
oscillations as seen in previous work\cite{branko}, as a result of
pair-breaking by the surface.

\subsection{Pair amplitude}

The above study of $\Delta(z)$ illustrates the detail,
and quality of the results. 
However,
since $\Delta(z)$ vanishes in the F side, this quantity cannot be used to
study superconducting proximity effects in the magnet. For this purpose
we now turn our attention to
the pair amplitude $F(z)$,
a quantity that  
directly reflects\cite{parks}
the superconducting
correlations in both F and S.
The main
panels in Figure \ref{pa1}, which repeat the arrangement of 
Fig. \ref{pp1},
show eight sets of results for $F(z)$, four for each of our two values
of $k_{FS}\xi_0$, for the
same  values of $I$ as in Fig.\ref{pp1}.
\begin{figure}[t]
{\epsfig{figure=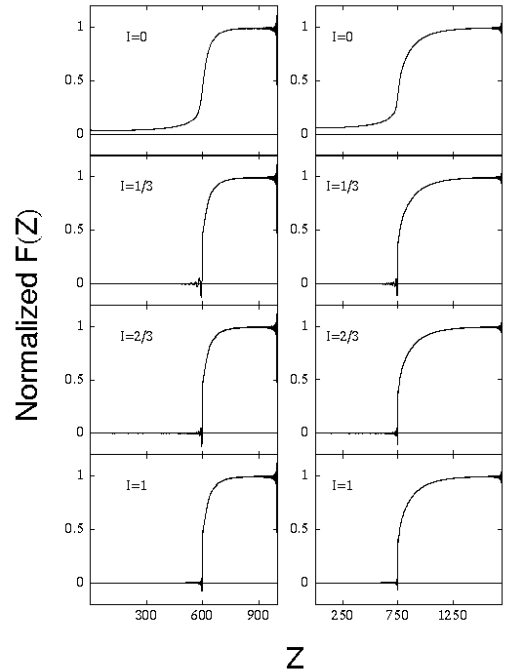,width=.45\textwidth}}
\newline
\caption{The  pair amplitude $F(z)$  (defined in Eq.(\ref{ca})), normalized
to its bulk value in the superconductor, plotted 
as a function of dimensionless distance $Z \equiv k_{FS} z$. 
Results are for the same coherence lengths
and exchange fields as in  Fig.\ref{pp1}, and with the
same panel arrangements.}
\label{pa1}
\end{figure} 
We have normalized $F(z)$ to its bulk value in the superconductor.
In the S region the curves  are  the same as
those for the corresponding $\Delta(z)$, seen above in Fig. \ref{pp1}.

Turning our attention to the
F side
of the interface
($z<d'$) in  Fig.\ref{pa1}, we first
look at the normal metal limit ($I=0$) in the top
panels.
We see that
as expected\cite{parks,falk} $F(z)$ decays only
extremely slowly in the normal region at $T=0$. 
In effect,
there is no mechanism to disrupt the Cooper pairs
from drifting across the interface\cite{falk}, therefore the
decay is very slow and occurs over
a length scale that is much
larger than $\xi_0$.
The most rapid change occurs near the interface,
where $F(z)$ decays very quickly before
flattening out.

In the remaining panels of Fig.\ref{pa1}
the effects of a finite exchange field are seen.
The situation is now very different and  $F(z)$ decays
to zero rather quickly
close to the interface,
with a slope that increases
with larger $I$. We will see below that the 
length that characterizes this fast decay
varies approximately as $1/I$. This
$1/I$ behavior was suggested
long ago\cite{parks} on the intuitive grounds 
that the exchange potentials seen by up and down
spin quasiparticles differ by $\pm I$, but this argument
has been criticized\cite{lg} as being merely an {\it ad hoc}
assumption. Our results show that the intuitive assumption
gives the  correct result. 
However, this fast decay is far from 
the whole story, as slightly away from
the interface a much slower 
oscillatory behavior can be seen
(note in particular the $I=1/3$ panel). 
This is {\it not} a finite size effect.
We have replotted this behavior in an expanded 
horizontal scale in Fig.\ref{inset}. 
\begin{figure}[t]
{\epsfig{figure=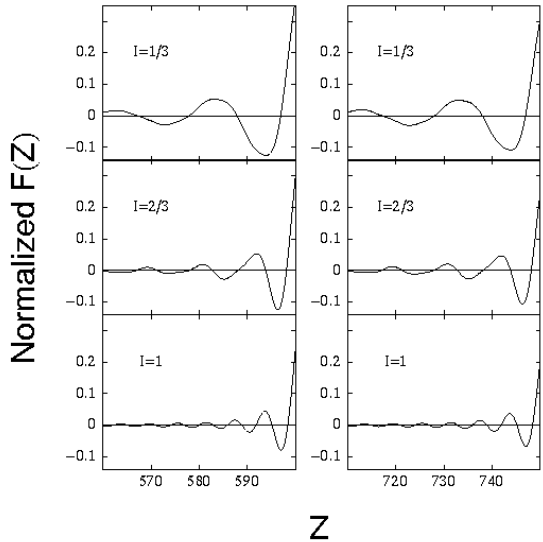,width=.45\textwidth}}
\newline
\caption{Detail of the behavior of the normalized pair amplitude
$F(z)$ near the interface, on the magnetic side. The six panels
shown correspond to the lower six panels of Fig. \ref{pa1}, but
the horizontal scale is expanded so that the oscillatory behavior
can be seen.}
\label{inset}
\end{figure} 
The  
wavelength 
of these oscillations 
clearly decreases with 
increased  $I$.  
Furthermore,  
the magnitude of $F(z)$ 
also attenuates with 
increasing $I$. This is in qualitative
agreement with past works employing
tight-binding\cite{ting} or
quasiclassical\cite{demler} methods.

Before we consider this behavior in more detail 
let us examine the $\xi_0$ dependence of these results for 
$F(z)$, by comparing the right and left column 
of Fig. \ref{pa1}. The spatial extent in which 
the changes in $F(z)$ take place 
is greater in the right column, since we are dealing now with  
a length 
scale given by a longer $\xi_0$. Apart 
from that,
the differences are hardly discernible, 
the only exception being the very slow decay for 
$I=0$, where the difference can be attributed to
the smaller value 
of $d'/\xi_0 \approx 4$ compared with $d'/\xi_0 \approx 12$ 
for the case of $k_{FS} \xi_0 = 50$. 
Thus we 
conclude that the role of $\xi_0$ is, in this range,
that of setting an overall scale. This should hold 
only when $\xi_0$  is much larger than the microscopic 
lengths in the problem and smaller than the geometrical dimensions.
It should break down in any other case. 
The exchange field tends to disrupt 
superconducting correlations over 
a length scale that is typically much smaller than 
$\xi_0$, so that the oscillations and 
characteristic decay of $F(z)$ in the magnetic
region are 
nearly independent of the $\xi_0$ considered
here. 
 
We are then led to conclude that  
when there is  an exchange field present,  
there are {\it two} phenomena to consider   
in describing
the spatial variations of $F(z)$ in  
the ferromagnet. 
The first is the  
short distance decay at the 
interface, to the point at which the pair amplitude first 
goes to zero. 
This is the region where $F(z)$ changes most 
rapidly. 
This decay can be characterized by a length scale which we will
denote by $\xi_1$, 
and 
define as  $\xi_1 \equiv d' - z_1$, where 
$z_1$ is the first point inside the magnet where
$F(z_1)=0$. 
The other important phenomenon
is the damped oscillations 
of $F(z)$ in the region $z<z_1$ (Fig. \ref{inset}). 
These oscillations cannot be fit to
an exponentially damped  form.
Instead, we find that in all cases
a much better fit to our results is afforded by the following 
expression:  
\begin{equation} 
F(z) =  
\alpha\frac{{\sin[ (z-d')/{\xi_2}]}}{(z-d')/{\xi_2}}, 
\label{xi2} 
\end{equation}   
where $\alpha$ is a constant, and 
the characteristic length $\xi_2$, which
in principle must be distinguished from $\xi_1$, can be extracted
from the results. 
Since the previously defined length, $\xi_1$, is small, 
the expression (\ref{xi2}) is valid for most 
of the ferromagnet region.  
To illustrate  
the range of its validity,  
in Fig.\ref{fit2} we give one example of a fit of the form  
Eq.(\ref{xi2}) to the pair amplitude. 
\begin{figure}[t]
{\epsfig{figure=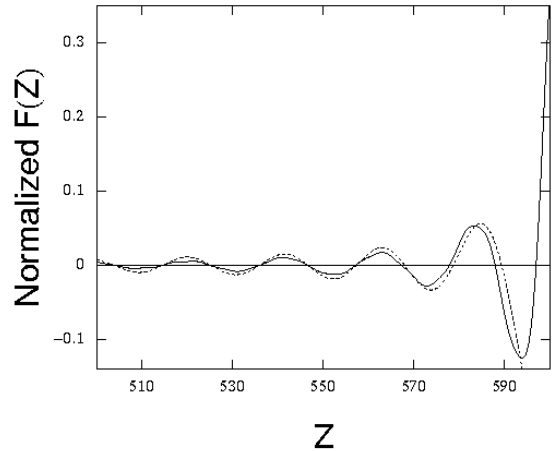,width=.45\textwidth}}
\newline
\caption{Example of a fit of the results for the
normalized $F(z)$ (solid curve) to the expression 
in Eq. (\ref{xi2}) (dashed line). The data displayed are for $k_{FS}\xi_0=50$
and $I=1/3$ (as shown in Figs. \ref{pa1}
and \ref{inset}}
\label{fit2}
\end{figure}  
We see that Eq.(\ref{xi2}) is an adequate 
fit 
for  the oscillatory  region,
however, within a distance $\xi_1$ of the 
interface, Eq.(\ref{xi2}) breaks down. 
At this point,  
$F(z)$  
rises upwards monotonically 
to match its value at the interface. 
In the spatial region where Eq.(\ref{xi2}) is 
valid, the quality of the fits deteriorates
somewhat 
for larger exchange fields ($0.4 \lesssim I <1$) 
because the  
spatial modulation of $F(z)$  
slightly deviates from 
the simple periodic sine curve 
given by  Eq.(\ref{xi2}). This small discrepancy 
can be glimpsed in the lower panels 
of Fig.\ref{inset}. 
The spatial structure becomes slightly nonperiodic, 
but overall  the functional form given by Eq.(\ref{xi2}) 
is still 
satisfactory.
 
The  oscillatory behavior 
of the pair amplitude as given  by (\ref{xi2}) 
is physically the result\cite{demler} of 
the exchange field, which  
creates 
electron and hole 
excitations 
in opposite spin bands. 
The pair amplitude 
involves  products 
of these particle and hole quasiparticle amplitudes (see Eq.(\ref{del2})). 
The superposition of these wavefunctions  
then creates oscillations  
on a length scale 
set by the difference 
between the  spin up and spin down wavevectors in  
the ferromagnet. One then expects\cite{demler}
a decay of the form (\ref{xi2}) with  
$\xi_2 \approx (k_{F\uparrow} - k_{F\downarrow})^{-1}$.
 We can then write: 
\begin{eqnarray} 
\label{xi2a} 
\xi_2 &\approx& 
\left[\sqrt{2m(E_{FM} + h_0)}-\sqrt{2m(E_{FM} - h_0)}\right]^{-1}
\nonumber \\ 
&=& k_{FS}^{-1} 
\frac{\sqrt{1+I}} 
{\sqrt{1+I} - \sqrt{1-I}}, 
\label{kdiff} 
\end{eqnarray} 
where in the last step we have used,
as previously mentioned, $E_{F\uparrow}/E_{FS}=1$. For 
small  $I$, Eq.(\ref{xi2a}) can be simplified to 
$k_{FS} \xi_2  
\approx  1/I$, 
showing  that 
$\xi_2$ is then
inversely proportional to the exchange  field.
At larger values of $I$ there are deviations, but these are small
since in the $I=1$ limit both Eq. (\ref{kdiff}) and the approximate
expression coincide. These oscillations are also related to those
responsible  for oscillatory coupling in structures involving magnetic
layers and superconducting spacers, 
\cite{sipr} and the
nonmonotonic behavior in the critical temperature, $T_c$, versus
the F-layer thickness in S/F/S junctions.\cite{radovic}
In particular, the  
sign change in the pair amplitude 
has the same physical origin as
the so called ``$\pi$ phase"
that exists in F/S multilayers,\cite{andreev2,baladie}
and the nonmonotonic variation of the Josephson current
with exchange field.\cite{prokic} 

Having introduced the two length 
scales $\xi_1$ and $\xi_2$ characterizing the 
superconducting proximity effect in the magnetic region, 
it is useful to compare their magnitude and behavior as functions
of $I$. 
The result of doing this is shown in 
Fig.\ref{fit}.  
\begin{figure}[t]
{\epsfig{figure=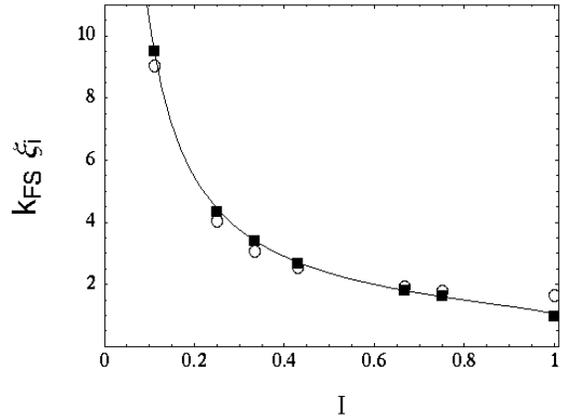,width=.45\textwidth}}
\newline
\caption{Exchange field, $I$,  dependence of the lengths $\xi_i,  
 i=1,2$, defined in the text. The circles are $\xi_1$, and 
the squares represent $\xi_2$.  
The curve is  
the expression in Eq.(\ref{kdiff}).
The results plotted are for $k_{FS}\xi_0=50$
but these quantities are nearly independent of $\xi_0$.}
\label{fit}
\end{figure}  
Data at additional values of $I$, not displayed in previous
Figures, is included.
For comparison, Eq.(\ref{xi2a}) is shown 
as the solid curve. We find that $\xi_2$ follows very
closely the expected theoretical expression, and that 
the other length, $\xi_1(I) \approx \xi_2(I)$. This is
because, as mentioned above, the expression\cite{parks}  $k_{FS}\xi_1=1/I$
nearly coincides numerically with the more complicated
result for $\xi_2$ as given above. Thus it turns out that the fast
decay and the spatial period of the oscillations are characterized
by lengths that are virtually identical.  
 
\subsection{Local density of states} 
To further investigate the F/S proximity effects, 
we focus now on another  experimentally accessible quantity, the  
local DOS.  
Advances in STM technology\cite{moussy}
have made it possible to perform
localized spectroscopic measurements with 
atomic scale resolution. 
We therefore present now 
the local DOS  
as a function of energy and position, as
calculated from Eq.(\ref{dos}) and the self-consistent spectra. 
All results below are normalized to the  
normal-state DOS in the S side and convolved with a Gaussian of 
width $0.01 \Delta_0$, to eliminate the
spectrum discretization resulting from the
finite size of the computational sample. We focus only on 
results for positive 
energies, since those for negative ones can be obtained by symmetry. 
We plot the results in terms of the normalized energy variable
$\varepsilon/\Delta_0$.
The locations chosen are given by the dimensionless
position measured relative to 
the interface in terms of the 
quantity $Z' \equiv k_{FS}(z-d')$. Thus, a positive
value of $Z'$ denotes a location within the superconductor.
 
We consider first the limit where the exchange field $I$ is zero.
In  Fig.\ref{dos1}, we show the DOS for four different
positions at each of the two values $k_{FS}\xi_0=50$ (left column)
and $k_{FS}\xi_0=200$ (right column).
\begin{figure}[t]
{\epsfig{figure=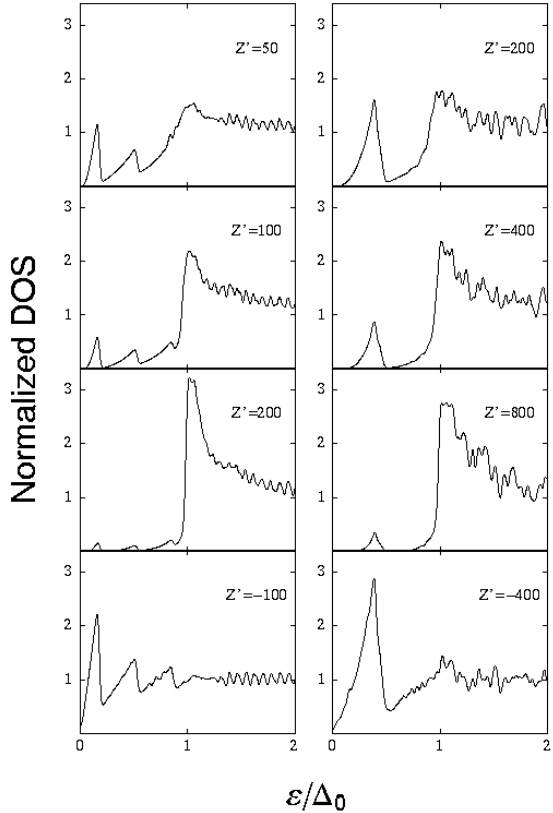,width=.45\textwidth}}
\newline
\caption{Normalized local DOS (see Eq.(\ref{dos})), plotted  
versus the dimensionless energy $\varepsilon/\Delta_0$ at $I=0$
for $k_{FS}\xi_0=50$ (left column) and $k_{FS}\xi_0=200$ (right column). 
The values of the dimensionless quantity 
$Z' \equiv k_{FS}(z-d')$  shown  by the labels are positions
relative to the 
interface. 
Each position displayed is a  multiple of the 
coherence length: from top to bottom 
the rows correspond to $Z'=\xi_0$, $Z'=2 \xi_0$, 
$Z'=4 \xi_0$, and $Z'= -2 \xi_0$. }
\label{dos1}
\end{figure}  
The three top rows in Fig.\ref{dos1} show the DOS 
on the S side. For the shorter coherence length 
results for the locations 
$Z'=50, 100$, and $200$ are shown. These are multiples
of the coherence length, and the same multiples are shown in the right column. 
Several pronounced peaks are visible inside the gap,  
due to a  finite number of bound states 
existing for $\varepsilon/ \Delta_0<1$. 
These states were predicted long ago
in a non self-consistent treatment by de Gennes and 
Saint-James\cite{degennes2}. 
These peaks diminish 
at greater distances inside the superconductor. 
On the corresponding panels in the right column,
we see that 
the number of de Gennes Saint-James peaks  
have been reduced. This is because
the number of bound states  
depends upon the 
coherence length  $\xi_0$, as well as 
on the superconductor and normal metal widths\cite{zaitlin}. 
In general, 
the number of such peaks decreases as 
$\xi_0/d'$ increases. The patterns seen at
$\varepsilon/\Delta_0>1$  are discussed below.

On the normal metal side, we see on
the bottom panels of Fig.\ref{dos1}, that  there is
no evidence of a gap, but 
a  pattern of jagged peaks appears 
in the DOS for $\varepsilon/\Delta_0 \lesssim 1$. At 
larger energies, interference patterns 
are seen, similar to those in the S side. At longer coherence lengths
this pattern is more coarse. This coarseness (which
is also seen in subsequent  figures) arises from
the finite value of   $N_\perp$. If this quantity is increased,
the pattern becomes smoother and more regular, as in the left column.
The remaining regular oscillations ultimately vanish as
$d$ and $d'$ tend to infinity.
We have chosen to display only one position in
the F side for $I=0$ 
since 
the overall behavior is nearly identical for all points 
in the normal metal. This is in agreement with
our observation in connection with the $I=0$ panels in 
Fig.\ref{pa1}, that the pair amplitude 
has a very slow rate of change. 
 
We now turn to the case of a  finite exchange field. 
Figure \ref{dos3} shows the DOS for $k_{FS} \xi_0 = 50$ (left column), and 
$k_{FS} \xi_0 = 200$ (right column) at $I=1/3$
for four positions very near the interface, within the magnetic material.
\begin{figure}[t]
{\epsfig{figure=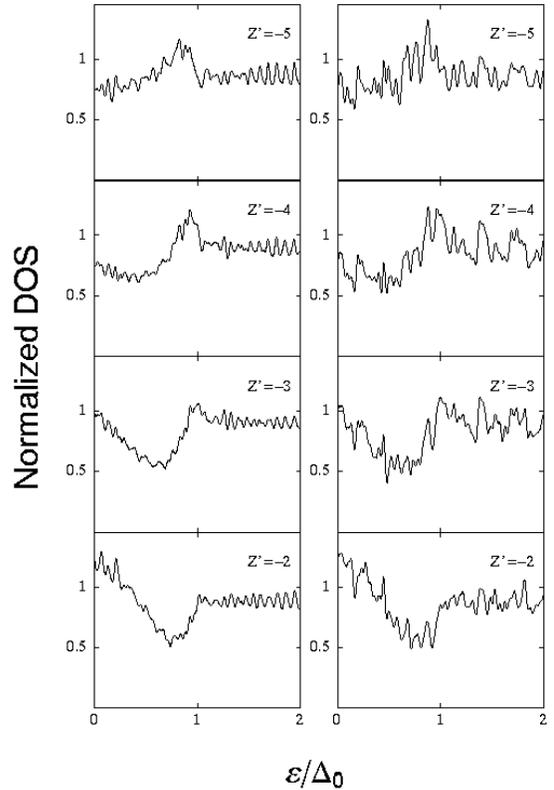,width=.45\textwidth}}
\newline
\caption{Normalized local DOS for  
$I=1/3$ at four positions in the ferromagnetic side, near the interface. 
The left column corresponds to  $k_{FS}\xi_0=50$
and the right one to $k_{FS}\xi_0=200$. 
$Z'$ is defined as in the previous Figure.}
\label{dos3}
\end{figure}  
This is done
to illustrate how changes in the local DOS with  
distance are correlated
with the rapid change in $F(z)$ near the interface.
Consider first the  distance $Z'=-5$ (top panels).  
This corresponds to the  location where
$F(z)$ has its more prominent minimum (see
Fig. \ref{inset}).
There is a weak minimum for the DOS at 
$\varepsilon/\Delta_0 = 0$, which is more prominent at the smaller $\xi_0$, 
and  with increasing energy
the DOS rises, 
until about $\varepsilon/\Delta_0 \approx 1$  
at which point a  peak occurs. 
For energies larger than $\Delta_0$ 
the DOS quickly settles down to 
its normal state value, unity in our
normalization. 
At $Z'=-4$, as $F(z)$ begins to rise,
we see, 
focusing on the range of 
energies less 
than $\Delta_0$, 
that 
the minimum of the DOS 
has begun to shift  away from zero. 
At $Z'=-3$,  
in the next row of panels,  
the DOS has now a marked minimum
at finite energies within the gap, 
at  $\varepsilon/\Delta_0 \approx 0.6$. 
The next position (last row) in Fig.\ref{dos3} 
shows a clear minimum of 
the DOS at  energies just below the gap. By comparing
the top and bottom rows of  
Fig.\ref{dos3}, we see that 
(for $\varepsilon/\Delta_0 \lesssim 1$) what were once 
dips and peaks in the DOS 
have now reversed roles. 
Fig.\ref{fit} shows that the length 
characterizing the fast rise of $F(z)$ is $k_{FS} \xi_1 \approx 3.5$
at $I=1/3$. 
The DOS   starts  the 
reversal process, as the interface is approached,
at around $Z'\approx -3.5$, as
seen in Fig. \ref{dos3}. The similarity between right
and left columns in this Figure reflects 
that the length scale $\xi_1$, defining the inversion 
point, is the same in both cases. 
The behavior of the DOS at larger values of $|Z'|$ is qualitatively similar, 
but as the oscillations die down
it becomes much less discernible.

We show also (see Fig.\ref{dos4}) the local DOS 
for three different positions in 
the superconductor side for the same $I=1/3$ and the 
shorter $\xi_0$. 
\begin{figure}[t]
{\epsfig{figure=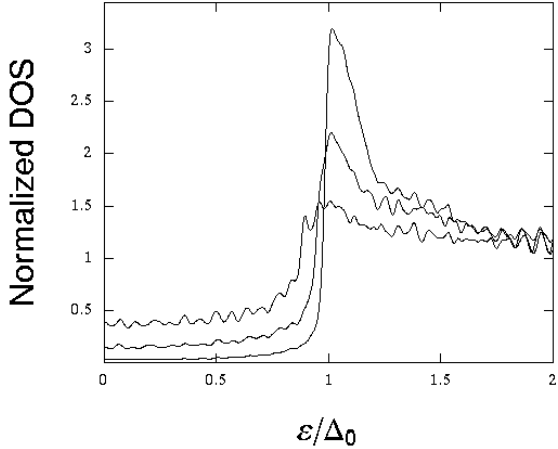,width=.45\textwidth}}
\newline
\caption{Local DOS  
at three positions inside the superconductor, for
$I=1/3$ and $k_{FS}\xi_0=50$. 
The  curves shown correspond to (from
top to bottom at small energy) $Z'=50$, $Z'=100$,
and $Z'=200$. These are the same values as in the top
three panels of the left column of Fig. \ref{dos1}.}
\label{dos4}
\end{figure}  
The de Gennes-Saint James peaks are 
now gone, with just a hint of small structure for $\varepsilon/ \Delta_0<1$ 
remaining at
$Z'=50$. 
This structure starts to become washed out 
at a distance of 
about $2 \xi_0$ from 
the interface. 
Finally, at $Z'=200$, the DOS is of  
the familiar BCS 
form, with a well defined gap and pronounced peak 
at $\varepsilon/\Delta_0 = 1$. 
In what follows, we focus only on 
the ferromagnetic region, since 
the overall behavior of the 
DOS in S at larger values of $I$ is quite similar to 
that seen in Fig.\ref{dos4}. 
 
In Fig.\ref{dos6}, we 
show the DOS for 
$I=2/3$.  
\begin{figure}[t]
{\epsfig{figure=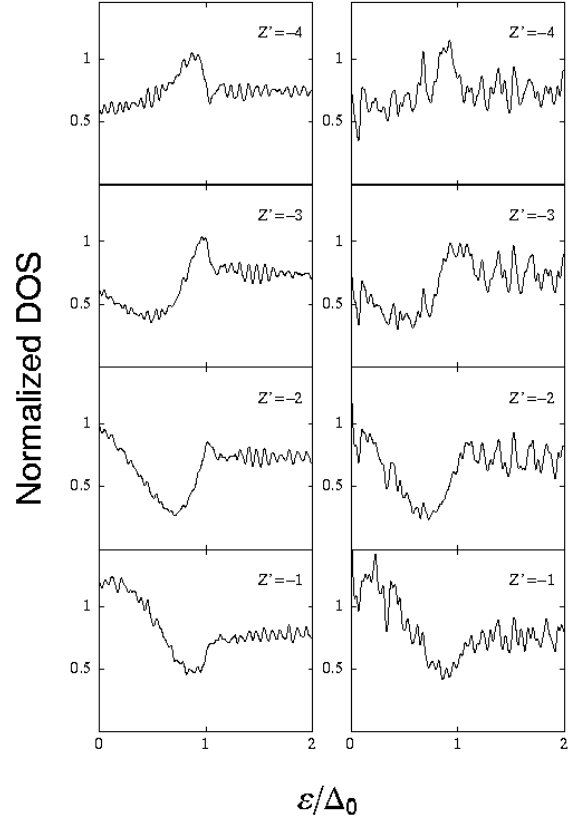,width=.45\textwidth}}
\newline
\caption{Normalized local DOS for $I=2/3$. The panel arrangement is the same
 as in Fig. \ref{dos3}.}
\label{dos6}
\end{figure}  
As in Fig.\ref{dos3}, we consider four 
spatial positions for each value of $\xi_0$, however the range is 
now closer 
to the interface, since the 
larger exchange field reduces the spatial extent of 
the superconducting correlations and the length $\xi_1$. 
Beginning at $Z'=-4$, Fig.\ref{dos6} (top) illustrates the formation 
of a small dip  
at low energies, and a continual rise up to  
$\varepsilon/\Delta_0 \approx 1$, after which 
we recover the bulk 
DOS limit for a ferromagnet with this polarization.  
With our normalization,
this value is smaller than unity. This is due to 
the decrease in the number of spin down states 
with increasing exchange field. 
In Fig.\ref{dos6} (second row), the minimum in the DOS 
has moved, while the peak still remains at $\varepsilon/\Delta_0 
\approx 1$.
At $Z'=-2$, (third row)  the DOS  
is already rising upwards at low energies.  
The previous dip 
in the DOS has  shifted to a  higher  energy, while 
a peak forms around zero energy. 
We again find consistency with the $\xi_1$ values given 
in Fig.\ref{fit}, where for $I=2/3$, $k_{FS} \xi_1 \approx 2.2$. 
Thus, as in the previous case, a reversal
of the DOS behavior occurs in the $\xi_1$ range.  
The bottom panel of Fig. \ref{dos6}
shows  the DOS at $Z'=-1$. We see that the zero energy
maximum has increased slightly from the previous row
and the minimum has shifted to energy $\Delta_0$. 
Again, this qualitative behavior is  independent of $\xi_0$
reflecting the independence of $\xi_1$ from $\xi_0$. 
Thus, we see here the same behavior we found for $I=1/3$
the only change being the different value of $\xi_1$.

We finally
consider  
in Fig.\ref{dos8} 
the DOS for a fully 
polarized (half metallic) ferromagnet ($I=1.0$). 
\begin{figure}[t]
{\epsfig{figure=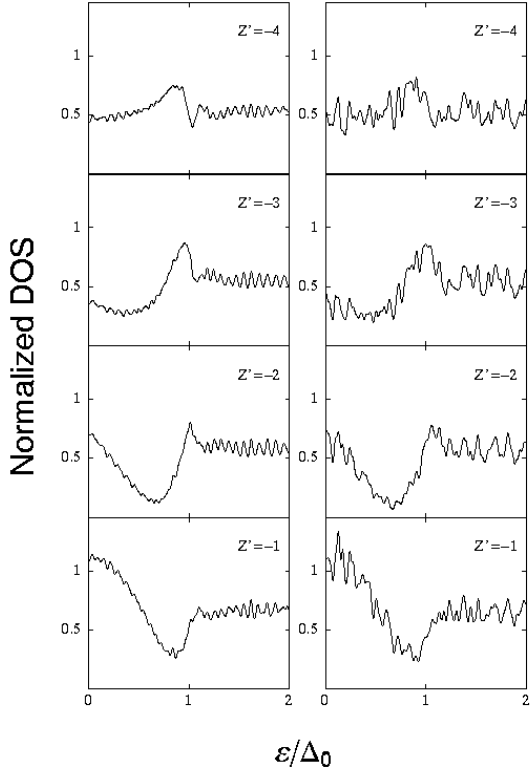,width=.45\textwidth}}
\newline
\caption{Normalized local DOS  for a fully polarized 
ferromagnet ($I=1.0$). 
Again, results for $k_{FS}\xi_0=50$ are 
in the left column, and those for 
 $k_{FS}\xi_0=200$ in the right column,
 and positions with respect to the interface are indicated.}
\label{dos8}
\end{figure}  
The locations $Z'$ are the same as in 
Fig.\ref{dos6}. 
The structure of the DOS at energies below the gap  for all positions 
has become smoother.
Because of the 
large exchange field, the 
reversal of the occupation 
of states occurs over a 
length scale $\xi_1$ which is now small (see Fig.\ref{inset}). 
Based on the previous fit 
in Fig.\ref{fit}, 
we find this point to be 
$Z' \approx 1.7$. 
Again, we find consistency between the 
pair amplitude and the DOS. 
Note that as one moves away from the interface,
the DOS tends to 1/2 at higher energies. This is due
to the total absence of the down band in this
half metallic limit. These remarks apply to both values of $\xi_0$. 

The above pertains to the proximity effect in the F side. We now  
investigate further whether the presence 
of the ferromagnet has any effect on 
the superconducting correlations. That any
effect is small is  shown already by 
Fig. \ref{pa1}. The influence of the magnet  
on S will be reflected in a nonzero value of the difference in the density of 
states for spin-up and spin-down electrons, $\delta N \equiv N_{\uparrow}- 
N_{\downarrow}$, where
$N_\uparrow$ and $N_\downarrow$
are the spin up and
spin down terms in Eq.(\ref{dos}) respectively. One might view this as
a self consistent determination
of an effective parameter $I(z)$ which
may extend into the superconductor.
We focus here only on the
case of a half metallic 
ferromagnet ($I=1$), and for illustration take $k_{FS}\xi_0=50$. 
Figure \ref{diff} shows that there is in fact a small
proximity effect into the superconductor, since  very close to the
interface the effective polarization is nonzero. 
\begin{figure}[t]
{\epsfig{figure=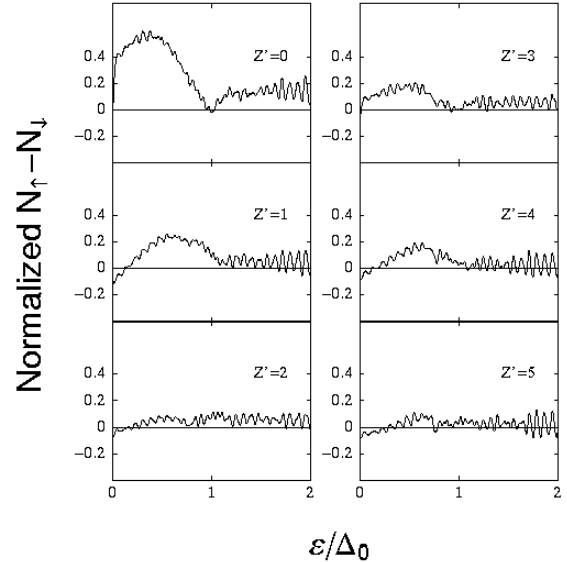,width=.45\textwidth}}
\newline
\caption{Leakage of magnetism into the superconductor: the 
quantity plotted difference between spin
up and spin down values of the local DOS, $\delta N \equiv N_{\uparrow}- 
N_{\downarrow}$, normalized as in previous Figures. 
This quantity is displayed at several positions 
just inside the superconductor. 
All results shown
in this Figure are for $k_{FS}\xi_0=50$. The values of
$Z'$ are indicated in the panels.}
\label{diff}
\end{figure}  
This effect 
is however short ranged, 
and we see that it nearly dies out before $Z'=5$. 
At very small exchange fields (of
order of the superconducting gap), we have 
found also a longer range proximity effect 
in the superconductor, similar to that found for dirty systems.\cite{fazio}

\section{conclusions} 
\label{conclusions}

We have introduced in this paper numerical techniques 
to accurately and self-consistently solve  the continuum BdG equations. 
We have shown how one can use these methods  
to perform a detailed study 
of F/S interfacial properties.
Our procedures allow
us to consider superconducting
and magnetic  proximity effects in a bulk system 
containing an F/S interface, even when the superconducting coherence
length is orders of magnitude larger than the interparticle distance.
In this work, we have used these techniques
to investigate the proximity effects 
for a clean F/S system. We have extracted the relevant
characteristic lengths through a careful analysis of the pair potential
and the pair amplitude, and we have shown how, near the interface, the behavior
of the pair amplitude correlates with that of the local DOS.
Our work extends well  beyond previous numerical computations
in the tight-binding case, valid only for very short values
of $\xi_0$, and  beyond 
theoretical work 
limited by 
quasiclassical approximations or restricted to regimes
where the mean free path or 
$\xi_0$ are very short.

On the S side we have found,
near the interface, a depletion
of both the pair potential and the pair amplitude.
This depletion extends over a length scale determined essentially by 
$\xi_0$, and hence nearly independent of the exchange field $I$. For the bulk
heterostructures we considered, the effect of varying $\xi_0$ 
in the range
studied ($k_{FS}\xi_0=50$ to
 $k_{FS}\xi_0=200$) was an effective rescaling of
the characteristic length that determines this depletion.
In the F region, for finite
values of the exchange field,
the pair amplitude exhibited 
a sharp monotonic decline near the interface, followed by
damped oscillations. 
The fast decay
was found to take place
over a length scale $\xi_1$  approximately inversely proportional to $I$, 
independent of the $\xi_0$, according the the
expression\cite{parks} $k_{FS}\xi_1 \approx 1/I$.
The oscillatory part of the
spatial variation of the pair amplitude could be  fit
to a simple sine function with an amplitude
decaying as
the inverse of distance from the interface.
We found that the  spatial period of the oscillations is
determined by the length difference 
$\xi_2 = (k_{F\uparrow} - k_{F\downarrow})^{-1}$
(the inverse of the difference between spin up and spin down
Fermi wavevectors)
provided that $I$ is not too large. 
This is in reasonable  agreement with 
previous theoretical expectations\cite{demler}. 
We have presented extensive results for the local DOS,
as a function of position and energy,
as obtained via the  self-consistent quasiparticle 
amplitudes and energies.
The periodic sign change in the pair amplitude
is found to be correlated with  oscillations in the local
DOS relative
to its normal state values. 
Finally, we 
verified also from the local DOS that
the effect of the
exchange field on superconducting correlations in S is minimal
(although nonzero):
the difference in the local DOS of spin up and spin down quasiparticles
vanishes except very close to the interface, at least
for $I\ge 1/3$.

Clearly,
the powerful methods and techniques for the self-consistent solution
of the BdG equations presented here open new vistas 
and possibilities for use in the study
of many other aspects of the F/S interface
and similar problems. 
A thorough investigation of the physical quantities and
characteristic lengths studied in this
paper,
incorporating other parameter regimes and 
the effects of finite temperature is needed,
and it can be straightforwardly carried out. Interface scattering, and 
superconductors with nodes in the pair potential 
(unconventional pair potentials), can be also easily considered.
Spin-flip
effects, and disorder in both F and S materials can also be incorporated.
By suitably changing the boundary conditions, self consistent solutions
of the tunneling spectroscopy problem in the long
$\xi_0$ regime will be obtainable.
Our numerical methods  are particularly suitable to the study
of mesoscopic structures involving 
F/S multilayers of differing 
thickness, where size effects may come into play.
The study of tunneling phenomena in  
non-equilibrium situations is also 
feasible 
by extension of our method to the time-dependent
BdG equations.

\acknowledgments

We thank P. Kraus and  A.M. Goldman for many conversations concerning
this problem. This work was supported in part by the Petroleum Research Fund,
administered by the ACS.

\end{twocolumn} 

\begin{references}
\bibitem{conver} In some cases it is more efficient to  
modify the result of the previous iteration  
before using it as the next input.
\end{references}
\end{document}